\documentclass[jkps,preprint,fleqn,showpacs,showkeys]{revtex4}
\usepackage{graphicx}
\usepackage{amssymb}
\usepackage{amsmath}
\usepackage{bm}
\begin{document}
\setcounter{page}{1}
\title{Matter Accretion by Brane-World Black Holes}
\author{Tiberiu \surname{Harko}}
\email{harko@hkucc.hku.hk}
\affiliation{Department of Physics, The University of Hong Kong,
Pok Fu Lam Road, Hong Kong, Hong Kong SAR, P. R. China}

\date[]{Received \today}

\begin{abstract}
The brane-world description of our universe entails a large extra
dimension and a fundamental scale of gravity that might be lower
by several orders of magnitude compared to the Planck scale. An
interesting consequence of the brane-world scenario is in the
nature of spherically symmetric vacuum solutions to the brane
gravitational field equations, with properties quite distinct as
compared to the standard black-hole solutions of general
relativity. We consider the spherically symmetric accretion of
matter onto brane-world black holes in terms of relativistic
hydrodynamics by assuming that the inflowing gas obeys a
polytropic equation of state. As a first step in this study, we
consider the accretion process in an arbitrary static, spherically
symmetric space-time, and show that the relativistic equations
require a transition to a supersonic flow in the solution. The
velocity, temperature, and density profiles are obtained for the
case of the polytropic equation of state. We apply the general
formalism to the study of the accretion properties of several classes
of brane-world black holes, and we obtain the distribution of the
main physical parameters of the gas. The astrophysical
determination of these physical quantities could discriminate, at
least in principle, between the different brane-world models, and
place some constraints on the existence of the extra dimensions.
\end{abstract}

\pacs{04.50.+h, 04.20.Jb, 04.20.Cv, 95.35.+d}

\keywords{Brane-world models; Accretion; Black holes}

\maketitle
\begin{center}
\section{INTRODUCTION}
\end{center}

The idea, proposed in Refs. \cite{RS99a} and \cite{RS99b},  that our
four-dimensional Universe might be a three-brane, embedded in a
five-dimensional space-time (the bulk) has attracted a
considerable interest in the past few years. According to the
brane-world scenario, the physical fields (electromagnetic,
Yang-Mills, etc.) in our four-dimensional Universe are confined to
the three brane. These fields are assumed to arise as fluctuations
of branes in string theories. Only gravity can freely propagate in
both the brane and bulk space-times, with the gravitational
self-couplings not being significantly modified. This model originated
from the study of a single $3$-brane embedded in five dimensions,
with the $5D$ metric given by $ds^{2}=e^{-f(y)}\eta _{\mu \nu
}dx^{\mu }dx^{\nu }+dy^{2}$, which, due to the appearance of the
warp factor, could produce a large hierarchy between the scale of
particle physics and gravity. Even if the fifth dimension is
uncompactified, standard $4D$ gravity is reproduced on the brane.
Hence, this model allows the presence of large, or even infinite,
non-compact extra dimensions. Our brane is identified to a domain
wall in a $5$-dimensional anti-de Sitter space-time.

Due to the correction terms coming from the extra dimensions,
significant deviations from the Einstein theory occur in brane-world models at very high energies \cite{SMS00,SSM00}. Gravity is
largely modified at the electro-weak scale of $1$ TeV. The
cosmological and astrophysical implications of the brane-world
theories have been extensively investigated in the
literature~\cite{all2,sol}.

Several classes of spherically symmetric solutions of the static
gravitational field equations in the vacuum on the brane have been
obtained~\cite{Ha03,Ma04,Ha04,Ha05}. As a possible physical
application of these solutions, the behavior of the angular velocity $%
v_{tg}$ of the test particles in stable circular orbits has been
considered~\cite{Ma04,Ha04,Ha05}. The general form of the
solution, together with two constants of integration, uniquely
determines the rotational velocity of the particle. In the limit
of large radial distances, and for a particular set of values of
the integration constants, the angular velocity tends to a constant
value. This behavior is typical for massive particles (hydrogen
clouds) outside galaxies and is usually explained by postulating
the existence of dark matter. The exact galactic metric, the
dark radiation, the dark pressure, and the lensing in the flat
rotation curves region in the brane-world scenario has been
obtained \cite{Ha05}.

For standard general relativistic spherical compact objects, the
exterior space-time is described by the Schwarzschild metric. In
the five dimensional brane-world models, the high-energy
corrections to the energy density, together with the Weyl stresses
from bulk gravitons, imply that on the brane, the exterior metric
of a static star is no longer the Schwarzschild metric
\cite{Da00}. The presence of the Weyl stresses also means that the
matching conditions do not have a unique solution on the brane;
the knowledge of the five-dimensional Weyl tensor is needed as a
minimum condition for uniqueness.

Static, spherically symmetric exterior vacuum solutions of the
brane-world models have been proposed first in Ref.~\cite{Da00} and in
Ref.~\cite{GeMa01}. The first of these solutions
\cite{Da00}, has the mathematical form of the Reissner-Nordstrom
solution of the standard general relativity, in which a tidal Weyl
parameter plays the role of the electric charge of the general
relativistic solution. The solution was obtained by imposing
the null energy condition on the 3-brane for a bulk having non-zero Weyl curvature, and it can be matched to the interior solution
corresponding to a constant-density brane-world star. A second
exterior solution, which also matches a constant density interior,
has been derived \cite{GeMa01}.

Two families of analytic solutions of the spherically symmetric
vacuum brane world model equations (with $g_{tt}\neq -1/g_{rr}$),
parameterized by the Arnowitt-Deser-Misner (ADM) mass and a Parameterized Post-Newtonian (PPN) parameter $\beta $ have
been obtained~\cite{cfm02}. Non-singular black-hole solutions
in the brane-world model have been considered by
relaxing the condition of the zero scalar curvature, but retaining
the null energy condition \cite{Da03}. The four-dimensional Gauss and Codazzi
equations for an arbitrary static spherically symmetric star in a
Randall--Sundrum type-II brane-world have been completely solved
on the brane in \cite{Vi03}. The on-brane boundary can be used to
determine the full $5$-dimensional space-time geometry. The
procedure can be generalized to solid objects, such as planets.

A method to extend into the bulk asymptotically-flat static
spherically symmetric brane-world metrics has been proposed
\cite{Ca03}. The exact integration of the field equations along
the fifth coordinate was done by using a multipole ($1/r$)
expansion. The results show that the shape of the horizon of the
brane-world black-hole solutions is very likely a flat ``pancake'' for
astrophysical sources.

The general solution to the trace of the 4-dimensional Einstein
equations for static, spherically-symmetric configurations has been
used as a basis for finding a general class of black-hole metrics,
containing one arbitrary function $g_{tt}=A(r)$, which vanishes at
some $r=r_h>0$ (the horizon radius) \cite{BMD03}. Under certain
reasonable restrictions, black-hole metrics are found, with or
without matter. Depending on the boundary conditions, the metrics
can be asymptotically flat or can have any other prescribed
asymptotic structure. For a review of the black hole properties and of the
lensing in the brane world models, see Ref.~\cite{MaMu05}.

It is generally expected that most of the astrophysical objects
grow substantially in mass via accretion. Recent observations
suggest that around most of the active galactic nuclei (AGN's) or
black hole candidates, there exist gas clouds surrounding the
central compact object, and an associated accretion disc, on a
variety of scales from a tenth of a parsec to a few hundred
parsecs \cite{UrPa95}. These clouds are assumed to form a
geometrically and optically thick torus (or warped disc), which
absorbs most of the ultraviolet radiation and the soft X-rays. The
gas exists in either a molecular or an atomic phase. The most
powerful evidence for the existence of super-massive black holes
comes from the VLBI imaging of molecular ${\rm H_2O}$ masers in
the active galaxy NGC 4258 \cite{Miyo95}. This imaging, produced
by Doppler shift measurements assuming Keplerian motion of the
masering source, has allowed a quite accurate estimate of the
mass of the central object, which has been found to be a $3.6\times 10^7M_{\odot
}$, a super massive dark object within $0.13$ parsecs. Hence,
important astrophysical information can be obtained from the
observation of the motion of the gas streams in the gravitational
field of compact objects.

The history of the theoretical study of the accretion of an ideal
fluid onto a compact object begins with Bondi's classic paper
\cite{Bo52}. A relativistic generalization of the Newtonian
accretion model was proposed by Michel \cite{Mi72} and further
considered in Refs.~\cite{Be78,BeGl81,Th81,ShTe83,Pa87, Pe88}.
In particular, Shapiro and Teukolsky \cite{ShTe83} gave a general
relativistic version of the Bondi \cite{Bo52} model, which is
known as the $(p-n)$ model. Another relativistic accretion model,
called the $(p-\rho )$ model, was proposed \cite{Ma99} and
further developed \cite{Ka06}.

It is not yet clear which equation of state is appropriate in the
description of the relativistic collapsing gas. There are two
commonly used polytropic equations of state: $p=K\rho ^{\Gamma }$
(the $(p-\rho )$ model \cite{ShTe83}) and $p=Cn^{\Gamma }$ (the
$(p-n)$ model \cite{Ma99}). Here $p$ is the pressure, $ \rho $ is
the density and $n$ is the baryonic mass density, while $K$ and
$C$ are constants. Numerical calculations show that the
predictions of the models are similar in most aspects. However, in
the ultra-relativistic regime the allowed band of the asymptotic
speed of sound and the mass accretion rate can be markedly
different \cite{Ma99}.

The determination of the accretion rate for an astrophysical
object can give strong evidence for the existence of a surface
for the object. A model in which Sgr A*, the $3.7\times 10^6
M_{\odot }$ super massive black hole candidate at the Galactic
center, may be a compact object with a thermally emitting surface
was considered \cite{BrNa06}. For very compact surfaces within
the photon orbit, the thermal assumption is likely to be a good
approximation because of the large number of rays that are
strongly gravitationally lensed back onto the surface. Given the
very low quiescent luminosity of Sgr A* in the near-infrared, the
existence of a hard surface, even in the limit in which the radius
approaches the horizon, places a severe constraint on the steady
mass accretion rate onto the source, ${\dot M}\le 10^{-12}
M_{\odot}$ yr$^{-1}$. This limit is well below the minimum
accretion rate needed to power the observed submillimeter
luminosity of Sgr A*, ${\dot M}\ge 10^{-10} M_{\odot}$ yr$^{}$.
Thus, from the determination of the accretion rate, it follows that
Sgr A* does not have a surface, that is, it must have an event
horizon. Therefore, the study of the accretion processes by compact
objects is a powerful indicator of their physical nature.

The stationary, spherically-symmetric accretion of dark energy
onto a Schwarzschild black hole was considered in terms of
relativistic hydrodynamics \cite{BDE05}. To model the dark
energy, the approximation of an ideal fluid was used. Constraints
on the gravastar models from accreting black holes were
obtained \cite{BrNa07}. In the study, two black hole candidates
known to have extraordinarily low luminosities, the super massive
black hole in the galactic center, Sagittarius A*, and the
stellar-mass black hole XTE J1118+480 have been used. The
observational results show that the length scale for modifications
of standard general relativity for the gravastar models must be
sub-Planckian.

It is the purpose of the present paper to study the matter
accretion by brane-world black holes. As a first step, we derive
the general equations for the spherically symmetric steady
accretion of a fluid in an arbitrary space-time. These equations
have a critical (sonic point), and only solutions passing through
it correspond to material falling into (or flowing out of) the
central accreting object, with monotonically increasing velocity
along the particle trajectory. The velocities at the sonic point
are obtained in a general form. In order to close the system of
equations of motion, we need to impose an equation of state
describing the thermodynamics properties of the inflowing matter.
By assuming that the equation of state is polytropic in the limit
of low temperature, we can solve the equations of motions,  and we can obtain the
velocity, density, and temperature profiles of the matter being
accreted by the black hole as functions of the
components of the metric tensor in a general form. This allows the velocity, temperature, and density of the matter
at the black hole's event horizon to be estimated.

By using the general formalism of accretion, we analyze the
accretion process for several brane-world black-hole solutions,
which have been previously obtained. Thus, we consider the
velocity, density, and temperature profiles of the inflowing gas,
the velocities at the sonic point, and the values of the velocity
and of the thermodynamical parameters at the event horizon for
three classes of vacuum solutions of the brane world models, which
have been obtained in Refs. \cite{Da00}, \cite{cfm02} and \cite{BMD03},
respectively. Due to the differences in the space-time metrics,
the velocity, density, and temperature profiles are different in
all these cases and different from the predictions of the
standard general relativistic Schwarzschild accretion model. Hence,
the study of the accretion processes may provide an effective
method to constrain the existence of the extra-dimensions and to
test the predictions of the brane-world models.

The present paper is organized as follows: We derive the general
equations for spherically symmetric steady accretion in Section
II. The case of a polytropic equation of state is analyzed in
Section III. The general formalism is applied to several classes
of brane-world black holes in Section IV. We discuss and conclude
our results in Section V.

\section{Steady spherical accretion in arbitrary spherically-symmetric spacetimes}

Let us consider the stationary, spherically symmetric accretion of
an ideal fluid in an arbitrary static, spherically symmetric
space-time, with a metric given by
\begin{equation}\label{metr}
ds^{2}=e^{\nu }c^{2}dt^{2}-e^{\lambda }dr^{2}-r^{2}\left( d\theta
^{2}+\sin ^{2}\theta d\varphi ^{2}\right) .
\end{equation}
Here $r$ is the radial coordinate, and $\theta $ and $\varphi $ are
the angular spherical coordinates, respectively. The metric tensor
components are assumed to be functions of the radial distance
only so that $\nu =\nu \left( r\right) $ and $\lambda =\lambda
\left( r\right) $. We model the fluid as an ideal fluid with the
energy-momentum tensor
\begin{equation}
T_{\mu \nu }=\left( \mu c^{2}+p\right) u_{\mu }u_{\nu }-pg_{\mu
\nu },
\end{equation}
where $\mu $ is the total energy density, $p$ is the pressure, and
$u^{\mu }=dx^{\mu }/ds$ is the four-velocity. Generally, the total
energy density can be represented as $\mu c^2=\rho c^2+\epsilon $,
where $\rho $ is the matter density and $\epsilon $ is the thermal
energy. The pressure is assumed to be an arbitrary function of the
density, $p=p\left( \rho \right) $. In the case of a radial flow,
the components of the four-velocity are $u^{0}=cdt/ds$,
$u_{0}=e^{\nu }u^{0}$, and $u^{1}=u=dr/ds$. The components of the
four-velocity are normalized so that $u_{\mu }u^{\mu }=1$ or
$u_{0}u^{0}+u_{1}u^{1}=e^{-\nu }u_{0}^{2}-e^{\lambda }u^{2}=1$.

The basic equations of motion of the fluid are the conservation of
the mass flux $J^{\mu }=\rho u^{\mu }$, $J_{;\mu }^{\mu }=0$, and
the conservation of the energy flux $T_{0;\mu }^{\mu }=0$, where
the semicolons denote the covariant derivative \cite{Mi72}. For a
steady spherically symmetric flow, the conservation equations have
the form
\begin{equation}
\frac{d}{dr}\left( \rho c^{2}ue^{\frac{\nu +\lambda
}{2}}r^{2}\right) =0,
\end{equation}
and
\begin{equation}
\frac{d}{dr}\left[ \left( \mu c^{2}+p\right) u_{0}ue^{\frac{\nu +\lambda }{2%
}}r^{2}\right] =0,
\end{equation}
respectively, giving
\begin{equation}
\rho c^{2}ue^{\frac{\nu +\lambda }{2}}r^{2}=C_{1},  \label{cons1}
\end{equation}
and
\begin{equation}
\left( \mu c^{2}+p\right) u_{0}ue^{\frac{\nu +\lambda
}{2}}r^{2}=C_{2}, \label{cons2}
\end{equation}
where $C_{1}$ and $C_{2}$ are constants of integration. By dividing Eq.~ (%
\ref{cons2}) by Eq.~(\ref{cons1}) and squaring gives
\begin{equation}
\left( \frac{\mu c^{2}+p}{\rho c^{2}}\right) ^{2}\left( e^{\nu
}+e^{\nu +\lambda }u^{2}\right) =C_{3},
\end{equation}
where $C_{3}=\left( C_{2}/C_{1}\right) ^{2}$.
By differentiating Eqs.~(\ref{cons1}) and (\ref{cons2}) and eliminating $%
d\ln \rho c^{2}$, we find
\begin{equation}
\left( r\frac{\nu ^{\prime }+\lambda ^{\prime }}{2}+2\right) \left\{ 2V^{2}-%
\frac{r\left[ \nu ^{\prime }e^{\nu }+\left( \nu ^{\prime }+\lambda
^{\prime }\right) e^{\nu +\lambda }u^2\right] }{\left( r\frac{\nu
^{\prime }+\lambda ^{\prime }}{2}+
2\right) \left( e^{\nu }+e^{\nu +\lambda }u^{2}\right) }%
\right\} \frac{dr}{r}+ 2\left[ V^{2}-\frac{e^{\nu +\lambda
}u^{2}}{e^{\nu }+e^{\nu +\lambda }u^{2}}\right] \frac{du}{u}=0,
\label{crit1}
\end{equation}
where we have denoted
\begin{equation}\label{defvel}
V^{2}=\frac{d\ln \left( \mu c^{2}+p\right) }{d\ln \rho c^{2}}-1.
\end{equation}

In the case of the Schwarzschild metric, we have $e^{\nu
}=e^{-\lambda }=1-2m/r$, with $m=GM/c^{2}$ being the total mass of the
accreting object. In this case, Eq.~(\ref{crit1}) reduces to the
basic equation for steady spherically symmetric accretion onto
compact objects, first derived in Ref.~\cite{Mi72}.

If one or the other of the bracketed factors in Eq.~(\ref{crit1})
vanishes, one has a turn-around point, and the solutions are
double-valued in either $r $ or $u$. Only solutions passing
through a critical point correspond to material falling into (or
flowing out of) the object with monotonically increasing velocity
along the particle trajectory. The critical point (also called the
sonic point) is located where all bracketed factors in Eq.
(\ref{crit1}) vanish \cite{Mi72}. Thus, in an arbitrary space-time,
the conditions for the existence of a critical point can be
formulated as
\begin{equation}\label{vel1}
2V^{2}\left( r\frac{\nu ^{\prime }+\lambda ^{\prime }}{2}+2\right) -\frac{r%
\left[ \nu ^{\prime }e^{\nu }+\left( \nu ^{\prime }+\lambda
^{\prime }\right) e^{\nu +\lambda }u^2\right] }{\left( e^{\nu
}+e^{\nu +\lambda }u^{2}\right) }=0,
\end{equation}
and
\begin{equation}\label{vel}
V^{2}-\frac{e^{\nu +\lambda }u^{2}}{e^{\nu }+e^{\nu +\lambda
}u^{2}}=0,
\end{equation}
respectively. In the case of the Schwarzschild metric, we obtain $%
u_{c}^{2}=m/2r_{c}$ and $V_{c}^{2}=u_{c}^{2}/\left(
1-3u_{c}^{2}\right) $, respectively \cite{Mi72}.
Substituting Eq.~(\ref{vel}) into Eq.~(\ref{vel1}) gives the
expression of the velocity at the sonic point as
\begin{equation}\label{uc}
u_{c}^{2}=\left. \frac{r\nu ^{\prime }e^{-\lambda }}{4}\right|
_{r=r_{c}}.
\end{equation}
Then, Eq.~(\ref{vel}) gives
\begin{equation}\label{vc}
V_{c}^{2}=\frac{e^{\lambda \left( r_{c}\right)
}u_{c}^{2}}{1+e^{\lambda \left( r_{c}\right) }u_{c}^{2}}.
\end{equation}

It is interesting to note that the tangential velocity of a particle
in a stable circular orbit in the space-time with the metric given by
Eq. (\ref{metr}) is given by $v_{tg}^{2}/c^2=r\nu ^{\prime }/2$
\citep{Ma04}. Therefore, we obtain the following relations for the
velocities at the sonic point and the tangential velocity of the
particles:
\begin{equation}
u_{c}^{2}=\left. \frac{v_{tg}^{2}e^{-\lambda }}{2c^{2}}\right|
_{r=r_{c}},
\end{equation}
and
\begin{equation}
V_{c}^{2}=\frac{v_{tg}^{2}\left( r_{c}\right)
/2c^{2}}{1+v_{tg}^{2}\left( r_{c}\right) /2c^{2}},
\end{equation}
respectively.

\begin{center}
\section{Accretion model with a polytropic equation of state}
\end{center}

In order to study the accretion processes in the brane-world
models, we need to specify the equation of state $p=p(\rho )$ of
the inflowing matter, which we assume to be in the form of a gas,
and the metric of the space-time. As for the equation of state, we
adopt the polytropic equation of state so that
\begin{equation}\label{eqs}
p=K\rho ^{\Gamma },
\end{equation}
with $K$ and $\Gamma $ being constants \cite{Mi72}. The temperature $T$
of the gas can be obtained from the ideal gas equation of state
$p=\rho k_{B}T/\mu m_{p}$, where $k_{B}$ is Boltzmann's constant,
$\mu $ is the mean molecular weight, and $m_{p}$ is the mass of the
proton. As a function of temperature, the pressure and the density
are given by
\begin{equation}\label{eqs1}
\rho =\left( \frac{c^{2}}{K}\right) ^{n}T_{p}^{n},p=K\left( \frac{c^{2}}{K}%
\right) ^{n+1}T_{p}^{n+1},
\end{equation}
where
\begin{equation}\label{eqs2}
n=\frac{1}{\Gamma -1},T_{p}=\frac{k_{B}T}{\mu
m_{p}c^{2}}=\frac{T}{\mu \times 1.09\times 10^{13}{\rm K}}.
\end{equation}

For the polytropic equation of state, we have \cite{Mi72}
\begin{equation}
p+\epsilon =(n+1)p.
\end{equation}
With the use of the equation of state of the gas, Eq.~(\ref
{defvel}) can be written as
\begin{equation}\label{velt}
V^{2}=\frac{(n+1)T_{p}}{n\left[ 1+(n+1)T_{p}\right] }.
\end{equation}
By estimating this equation at the sonic point, where the gas
temperature is $T_{pc}$, and comparing with Eq.~(\ref{vc}) gives
\begin{equation}\label{tpc}
T_{pc}=\frac{n}{n+1}\frac{e^{\lambda \left( r_{c}\right) }u_{c}^{2}}{%
1+(1-n)e^{\lambda \left( r_{c}\right) }u_{c}^{2}}.
\end{equation}

The equations of motion of the inflowing gas with a polytropic
equation of state are given by
\begin{equation}
c^{2}\left( \frac{c^{2}}{K}\right) ^{n}T_{p}^{n}ur^{2}e^{\left(
\nu +\lambda \right) /2}=B_{1},
\end{equation}
and
\begin{equation}
\left[ 1+\left( n+1\right) T_{p}\right] ^{2}\left( e^{\nu }+e^{\nu
+\lambda }u^{2}\right) =B_{3},
\end{equation}
respectively, with $B_{1}$ and $B_{3}$ being constants of integration.
Evaluating Eqs. (\ref{bl1}) and (\ref{bl2}) at $r=r_{\infty }$,
where $e^{\nu \left( r_{\infty }\right) }=e^{\nu _{\infty }}$,
$e^{\lambda \left( r_{\infty
}\right) }=e^{\lambda _{\infty }}$ and the velocity and the temperature are $%
u\left( r_{\infty }\right) =u_{\infty }$ and $T_{p}\left(
r_{\infty }\right) =T_{p\infty }$, respectively,  gives
\begin{equation}
B_{1}=c^{2}\left( \frac{c^{2}}{K}\right) ^{n}T_{p\infty
}^{n}u_{\infty }r_{\infty }^{2}e^{\left( \nu _{\infty }+\lambda
_{\infty }\right) /2},
\end{equation}
and
\begin{equation}
B_{3}=\left[ 1+\left( n+1\right) T_{p\infty }\right] ^{2}\left(
e^{\nu _{\infty }}+e^{\nu _{\infty }+\lambda _{\infty }}u_{\infty
}^{2}\right) ,
\end{equation}
respectively.

Therefore, the two equations in the unknowns $T_{p}$ and $u$
describing the motion of the polytropic gas in an arbitrary
static, spherically symmetric metric are
\begin{equation}\label{eqnf1}
T_{p}^{n}ur^{2}e^{\left( \nu +\lambda \right) /2}=T_{p\infty
}^{n}u_{\infty }r_{\infty }^{2}e^{\left( \nu _{\infty }+\lambda
_{\infty }\right) /2},
\end{equation}
and
\begin{equation}\label{eqnf2}
\left[ 1+\left( n+1\right) T_{p}\right] ^{2}\left( e^{\nu }+e^{\nu
+\lambda }u^{2}\right) = \left[ 1+\left( n+1\right) T_{p\infty
}\right] ^{2}\left( 1+e^{\nu _{\infty }+\lambda _{\infty
}}u_{\infty }^{2}\right) .
\end{equation}
Evaluating Eq. (\ref{eqnf1}) at the sonic point $r=r_c$ gives
an algebraic equation for the determination of $r_c$,
\begin{equation}
T_{pc}^{n}r_{c}^{5/2}e^{\nu \left( r_{c}\right) }\sqrt{\nu ^{\prime }}%
\left.\right|_{r=r_{c}}=2T_{p\infty }^{n}u_{\infty }r_{\infty
}^{2}e^{\left( \nu _{\infty }+\lambda _{\infty }\right) /2}.
\end{equation}
By assuming that $T_{p}<<1$ and $e^{\nu _{\infty }+\lambda
_{\infty }}u_{\infty }^{2}<<1$, Eq. (\ref{eqnf2}) immediately
gives
\begin{equation}
u\left( r\right) \approx \left( 1-e^{\nu }\right) ^{1/2}e^{-\left(
\nu +\lambda \right) /2}.
\end{equation}
By substituting this expression of the velocity in Eq.
(\ref{eqnf1}), we obtain the temperature and the density profiles of
the cold accreting gas as
\begin{equation}
T_{p}(r)\approx T_{p\infty }\frac{\left[ u_{\infty }r_{\infty
}^{2}e^{\left( \nu _{\infty }+\lambda _{\infty }\right) /2}\right]
^{1/n}}{r^{2/n}\left( 1-e^{\nu }\right) ^{1/2n}},
\end{equation}
and
\begin{equation}
\rho \left( r\right) \approx \left( \frac{c^{2}}{K}\right)
\frac{u_{\infty
}r_{\infty }^{2}e^{\left( \nu _{\infty }+\lambda _{\infty }\right) /2}}{%
r^{2}\left( 1-e^{\nu }\right) ^{1/2}},
\end{equation}
respectively. At the event horizon of the black hole, $r=r_{h}$. Therefore, we
obtain the velocity $u_h=u\left(r_h\right)$, the temperature
$T_{ph}=T_{p}\left( r_{h}\right) $ and the density $\rho _{h}=\rho
\left( r_{h}\right) $ of the gas at the event horizon of the black
hole as
\begin{equation}
u_{h}\approx \left[ 1-e^{\nu \left( r_{h}\right) }\right]
^{1/2}e^{-\left[ \nu \left( r_{h}\right) +\lambda \left(
r_{h}\right) \right] /2},
\end{equation}
\begin{equation}
T_{ph}\approx T_{p\infty }\frac{\left[ u_{\infty }r_{\infty
}^{2}e^{\left( \nu _{\infty }+\lambda _{\infty }\right) /2}\right]
^{1/n}}{r_{h}^{2/n}\left[ 1-e^{\nu \left( r_{h}\right) }\right]
^{1/2n}},
\end{equation}
and
\begin{equation}
\rho _{h}\approx \left( \frac{c^{2}}{K}\right) \frac{u_{\infty
}r_{\infty }^{2}e^{\left( \nu _{\infty }+\lambda _{\infty }\right)
/2}}{r_{h}^{2}\left[ 1-e^{\nu \left( r_{h}\right) }\right]
^{1/2}},
\end{equation}
respectively.

If $\exp \left[ \nu \left( r_{h}\right) +\lambda \left(
r_{h}\right) \right] =1$, as is the case for the standard
Schwarzschild solution of general relativity, and because at the
event horizon $\exp \left[ \nu \left( r_{h}\right) \right] =0$,
the gas particles reach the ''surface'' (event horizon) of the
black hole with a four-velocity equal to the speed of light,
$u_h\approx 1$. However, for black holes for which the event
horizon is located so that $\exp %
\left[ \nu \left( r_{h}\right) +\lambda \left( r_{h}\right)
\right] \neq 1$, $\exp \left[ -\lambda \left( r_{h}\right) \right]
=0$, and $\exp \left[ \nu \left( r_{h}\right) \right] \neq 0$, we
have $u_{h}\approx 0$. Such black holes occur is the brane world
models.

An important physical quantity in the description of the accretion
is the speed of sound $a^{2}=\partial p/\partial \rho =\Gamma
p/\rho =c^{2}\Gamma T_{p}$. Hence, the speed of sound at infinity
is related to the temperature at infinity by the simple relation
$T_{p\infty }=a_{\infty }^2/\Gamma c^2$. At the sonic point, the
speed of sound is
\begin{equation}\label{ac}
a_{c}=c\sqrt{\frac{e^{\lambda \left( r_{c}\right) }u_{c}^{2}}{%
1+(1-n)e^{\lambda \left( r_{c}\right) }u_{c}^{2}}}\leq c.
\end{equation}

Sometimes an alternative form of the equation of state of the gas
is used,
by assuming that the pressure $p$ is related to the baryon number density $%
n_{B}$ by the polytropic relation $p=Kn_{B}^{\Gamma }$. $n_{B}$
and $\rho $ can be related by the general relation $n_{B}=\exp
\left[ \int d\rho c^{2}/\left( \rho c^{2}+K\rho ^{\Gamma }\right)
\right] $, which can be integrated to give $n_{B}=\rho \left(
c^{2}/K\right) ^{-n}\left( 1+a^{2}/\Gamma c^{2}\right) ^{-n}$.
Therefore, in the present model, the baryon number density is given
as a function of the radial distance $r$ by
\begin{equation}
n_{B}\approx \left( \frac{c^{2}}{K}\right) ^{1-n}\frac{u_{\infty
}r_{\infty }^{2}e^{\left( \nu _{\infty }+\lambda _{\infty }\right)
/2}}{r^{2}\left( 1-e^{\nu }\right) ^{1/2}\left( 1+a^{2}/\Gamma
c^{2}\right) ^{n}}.
\end{equation}

\section{Gravitational field equations in the brane-world models}\label{field}

In the present section, we briefly describe the basic mathematical
formalism of the brane-world models, and we present the
spherically symmetric static vacuum field equations. The solutions
of the vacuum field equations on the brane physically describe the
brane-world black holes.

\subsection{Gravitational Field Equations on the Brane}

We start by considering a five-dimensional (5D) spacetime (the
bulk) with a single four-dimensional (4D) brane, on which matter
is confined. The 4D brane world $({}^{(4)}M,g_{\mu \nu })$ is
located at a hypersurface $\left(
B\left( X^{A}\right) =0\right) $ in the 5D bulk spacetime $%
({}^{(5)}M,g_{AB}) $, whose coordinates are described by $%
X^{A},A=0,1,...,4$. The induced 4D coordinates on the brane are
$x^{\mu },\mu =0,1,2,3$.
The action of the system is given by $S=S_{bulk}+S_{brane}$, where
\begin{equation}
S_{bulk}=\int_{{}^{(5)}M}\sqrt{-{}^{(5)}g}\left[ \frac{1}{2k_{5}^{2}}{}%
^{(5)}R+{}^{(5)}L_{m}+\Lambda _{5}\right] d^{5}X,
\end{equation}
and
\begin{equation}
S_{brane}=\int_{{}^{(4)}M}\sqrt{-{}^{(5)}g}\left[
\frac{1}{k_{5}^{2}}K^{\pm }+L_{brane}\left( g_{\alpha \beta },\psi
\right) +\lambda _{b}\right] d^{4}x,
\end{equation}
where $k_{5}^{2}=8\pi G_{5}$ is the 5D gravitational constant,
${}^{(5)}R$ and ${}^{(5)}L_{m}$ are the 5D scalar curvature and
the matter Lagrangian in the bulk, $L_{brane}\left( g_{\alpha
\beta },\psi \right) $ is the 4D
Lagrangian, which is given by a generic functional of the brane metric $%
g_{\alpha \beta }$ and of the matter fields $\psi $, $K^{\pm }$ is
the trace of the extrinsic curvature on either side of the brane,
and $\Lambda _{5}$ and $\lambda _{b}$ (the constant brane tension)
are the negative vacuum energy densities in the bulk and on the
brane, respectively~\cite{SMS00}.

The Einstein field equations in the bulk are given by~\cite{SMS00}
\begin{equation}
{}^{(5)}G_{IJ}=k_{5}^{2} {}^{(5)}T_{IJ},\qquad
{}^{(5)}T_{IJ}=-\Lambda _{5}
{}^{(5)}g_{IJ}+\delta(B)\left[-\lambda_{b}
{}^{(5)}g_{IJ}+T_{IJ}\right] ,
\end{equation}
where ${}^{(5)}T_{IJ}\equiv - 2\delta {}^{(5)}L_{m}/\delta
{}^{(5)}g^{IJ} +{}^{(5)}g_{IJ} {}^{(5)}L_{m}$ is the
energy-momentum tensor of bulk matter fields while $T_{\mu \nu }$
is the energy-momentum tensor localized on the brane, which is
defined by $T_{\mu \nu }\equiv -2\delta L_{brane}/\delta g^{\mu
\nu }+g_{\mu \nu }\text{ }L_{brane}$.
The delta function $\delta \left( B\right) $ denotes the
localization of the
brane contribution. In the 5D spacetime, a brane is a fixed point of the $%
Z_{2}$ symmetry. The basic equations on the brane are obtained by
projections onto the brane world. The induced 4D metric is $%
g_{IJ}={}^{(5)}g_{IJ}-n_{I}n_{J}$, where $n_{I}$ is the space-like
unit vector field normal to the brane hypersurface ${}^{(4)}M$. In
the following we assume ${}^{(5)}L_{m}=0$. In the brane-world
models only gravity can probe the extra dimensions.

Assuming a metric of the form
$ds^{2}=(n_{I}n_{J}+g_{IJ})dx^{I}dx^{J}$, with $n_{I}dx^{I}=d\chi
$ being the unit normal to the $\chi =\mathrm{constant}$ hypersurfaces
and $g_{IJ}$ being the induced metric on $\chi =\mathrm{constant}$
hypersurfaces, the effective 4D gravitational equation on the
brane takes the form~\cite{SMS00}:
\begin{equation}
G_{\mu \nu }=-\Lambda g_{\mu \nu }+k_{4}^{2}T_{\mu \nu
}+k_{5}^{4}S_{\mu \nu }-E_{\mu \nu },  \label{Ein}
\end{equation}
where $S_{\mu \nu }$ is the local quadratic energy-momentum
correction
\begin{equation}
S_{\mu \nu }=\frac{1}{12}TT_{\mu \nu }-\frac{1}{4}T_{\mu
}{}^{\alpha }T_{\nu \alpha }+\frac{1}{24}g_{\mu \nu }\left(
3T^{\alpha \beta }T_{\alpha \beta }-T^{2}\right) ,
\end{equation}
and $E_{\mu \nu }$ is the non-local effect from the free bulk
gravitational
field, the transmitted projection of the bulk Weyl tensor $C_{IAJB}$, $%
E_{IJ}=C_{IAJB}n^{A}n^{B}$, with the property $E_{IJ}\rightarrow
E_{\mu \nu }\delta _{I}^{\mu }\delta _{J}^{\nu }\quad $as$\quad
\chi \rightarrow 0$. We have also denoted $k_{4}^{2}=8\pi G$, with
$G$ being the usual 4D gravitational constant.
The 4D cosmological constant, $\Lambda $, and the 4D coupling constant, $%
k_{4}$, are related by $\Lambda =k_{5}^{2}(\Lambda
_{5}+k_{5}^{2}\lambda _{b}^{2}/6)/2$ and
$k_{4}^{2}=k_{5}^{4}\lambda _{b}/6$, respectively. In the limit
$\lambda _{b}^{-1}\rightarrow 0$ we recover standard general
relativity \cite{SMS00}.

The Einstein equation in the bulk and the Codazzi equation also
imply
conservation of the energy-momentum tensor of the matter on the brane, $%
D_{\nu }T_{\mu }{}^{\nu }=0$, where $D_{\nu }$ denotes the brane
covariant derivative. Moreover, from the contracted Bianchi
identities on the brane, it follows that the projected Weyl tensor
obeys the constraint $D_{\nu }E_{\mu }{}^{\nu }=k_{5}^{4}D_{\nu
}S_{\mu }{}^{\nu }$.

The symmetry properties of $E_{\mu \nu }$ imply that, in general, we
can
decompose it irreducibly with respect to a chosen $4$-velocity field $%
u^{\mu} $ as $E_{\mu \nu }=-k^{4}\left[ U\left( u_{\mu }u_{\nu}
+\frac{1}{3}h_{\mu \nu }\right) +P_{\mu \nu }+2Q_{(\mu
}u_{\nu)}\right]$, where $k=k_{5}/k_{4}$, $h_{\mu \nu }=g_{\mu \nu
}+u_{\mu }u_{\nu }$ projects orthogonal to $u^{\mu }$, the ``dark
radiation'' term $U=-k^{-4}E_{\mu \nu }u^{\mu }u^{\nu }$ is a
scalar, $Q_{\mu }=k^{-4}h_{\mu }^{\alpha }E_{\alpha \beta
}u^{\beta }$ is a spatial vector and $P_{\mu \nu }=-k^{-4}\left[
h_{(\mu }\text{ }^{\alpha }h_{\nu )}\text{ }^{\beta
}-\frac{1}{3}h_{\mu \nu }h^{\alpha \beta }\right] E_{\alpha \beta
}$ is a spatially, symmetric and trace-free tensor~\cite{Da00}.

In the case of the vacuum state, we have $\rho =p=0$ and $T_{\mu \nu
}\equiv 0$, consequently, $S_{\mu \nu }\equiv 0$. Therefore, the
field equation describing a static brane takes the form
\begin{equation}
R_{\mu \nu }=-E_{\mu \nu }+\Lambda g_{\mu \nu },
\end{equation}
with the trace $R$ of the Ricci tensor $R_{\mu \nu }$ satisfying
the condition $R=R_{\mu }^{\mu }=4\Lambda $.

In the vacuum case $E_{\mu \nu }$ satisfies the constraint $D_{\nu
}E_{\mu
}{}^{\nu }=0$. In an inertial frame at any point on the brane, we have $%
u^{\mu }=\delta _{0}^{\mu }$ and $h_{\mu \nu
}=\mathrm{diag}(0,1,1,1)$. In a static vacuum, $Q_{\mu }=0$, and the
constraint for $E_{\mu \nu }$ takes the form~ \cite{GeMa01}
\begin{equation}
\frac{1}{3}D_{\mu }U+\frac{4}{3}UA_{\mu }+D^{\nu }P_{\mu \nu
}+A^{\nu }P_{\mu \nu }=0,
\end{equation}
where $A_{\mu }=u^{\nu }D_{\nu }u_{\mu }$ is the 4-acceleration.
In the
static spherically symmetric case, we may chose $A_{\mu }=A(r)r_{\mu }$ and $%
P_{\mu \nu }=P(r)\left( r_{\mu }r_{\nu }-\frac{1}{3}h_{\mu \nu
}\right) $, where $A(r)$ and $P(r)$ (the ``dark pressure'') are
some scalar functions of the radial distance~$r$, and~$r_{\mu }$
is a unit radial vector~\cite{Da00}.

\subsection{Brane-world Black Holes}

In the following, we will restrict our study to the static and
spherically symmetric metric given by
\begin{equation}
ds^{2}=-e^{\nu (r)}dt^{2}+e^{\lambda (r)}dr^{2}+r^{2}\left(d\theta
^{2}+\sin ^{2}\theta d\phi ^{2}\right). \label{metr1}
\end{equation}
With the metric given by Eq.~(\ref{metr1}), the gravitational field
equations and the effective energy-momentum tensor conservation
equation in the vacuum take the forms~\cite{Ha03,Ma04}
\begin{equation}
-e^{-\lambda }\left( \frac{1}{r^{2}}-\frac{\lambda ^{\prime }}{r}\right) +%
\frac{1}{r^{2}}=3\alpha U+\Lambda ,  \label{f1}
\end{equation}
\begin{equation}
e^{-\lambda }\left( \frac{\nu ^{\prime }}{r}+\frac{1}{r^{2}}\right) -\frac{1%
}{r^{2}}=\alpha \left( U+2P\right) -\Lambda , \label{f2}
\end{equation}
\begin{equation}
\frac{1}{2}e^{-\lambda }\left( \nu ^{\prime \prime }+\frac{\nu ^{\prime 2}}{2%
}+\frac{\nu ^{\prime }-\lambda ^{\prime }}{r}-\frac{\nu ^{\prime
}\lambda ^{\prime }}{2}\right) =\alpha \left( U-P\right) -\Lambda
,  \label{f3}
\end{equation}
\begin{equation}
\nu ^{\prime }=-\frac{U^{\prime }+2P^{\prime
}}{2U+P}-\frac{6P}{r\left( 2U+P\right) },  \label{f4}
\end{equation}
where $^{\prime }=d/dr$, and we have denoted $\alpha =16\pi
G/k^{4}\lambda _{b}$.

The field equations, Eqs.~(\ref{f1})--(\ref{f3}), can be interpreted as
describing an anisotropic ''matter distribution,'' with the
effective energy density $\rho ^{\mathrm{eff}}$, radial pressure
$P^{\mathrm{eff}}$, and orthogonal pressure $P_{\perp
}^{\mathrm{eff}}$, respectively, so that $\rho
^{\mathrm{eff}}=3\alpha U+\Lambda $, $P^{\mathrm{eff}}=\alpha
U+2\alpha P-\Lambda $, and $P_{\perp }^{\mathrm{eff}}=\alpha
U-\alpha P-\Lambda $, respectively, which gives the condition
$\rho ^{\mathrm{eff}}-P^{\mathrm{eff}}-2P_{\perp }^{\mathrm{eff}}=4\Lambda =%
\mathrm{constant}$. This is expected for the `radiation'-like
source, for which the projection of the bulk Weyl tensor is
trace-less, $E_{\mu }^{\mu }=0$.

Equation~(\ref{f1}) can immediately be integrated to give
\begin{equation}
e^{-\lambda }=1-\frac{C_{1}}{r}-\frac{GM_{U}\left( r\right) }{r}-\frac{%
\Lambda }{3}r^{2},  \label{m1}
\end{equation}
where $C_{1}$ is an arbitrary constant of integration, and we
have used $GM_{U}\left( r\right) =3\alpha \int_{0}^{r}U(r)r^{2}dr$.
The function $M_U$ is the gravitational mass corresponding to the
dark radiation term (the dark mass). For $U=0$, the metric
coefficient given by Eq.~(\ref{m1}) must tend to the standard
general relativistic Schwarzschild metric coefficient, which gives
$C_{1}=2GM$, where $M = \mathrm{constant}$ is the baryonic (usual)
mass of a gravitating system.

By substituting $\nu ^{\prime }$ given by Eq.~(\ref{f4}) into
Eq.~(\ref{f2}) and using Eq.~(\ref{m1}), we obtain the
following system of differential equations satisfied by the dark
radiation term $U$, the dark pressure $P$, and the dark mass
$M_{U}$, describing the vacuum gravitational field exterior to a
massive body, in the brane-world model \cite{Ha03}:
\begin{equation}
\frac{dM_{U}}{dr}=\frac{3\alpha }{G}r^{2}U.  \label{e2}
\end{equation}
\begin{equation}
\frac{dU}{dr}=-\frac{\left( 2U+P\right) \left[
2GM+GM_{U}-\frac{2}{3}\Lambda
r^{3}+\alpha \left( U+2P\right) r^{3}\right] }{r^{2}\left( 1-\frac{2GM}{r}-%
\frac{M_{U}}{r}-\frac{\Lambda }{3}r^{2}\right)
}-2\frac{dP}{dr}-\frac{6P}{r}, \label{e1}
\end{equation}
To close the system, a supplementary functional relation between
one of the unknowns, $U$, $P$ or $M_{U}$, is needed. Generally,
this equation of state
is given in the form $P=P(U)$. Once this relation is known, Eqs.~(\ref{e2}%
)--(\ref{e1}) give a full description of the geometrical
properties of the vacuum on the brane.

In the following we will restrict our analysis to the case
$\Lambda =0$. Then, the system of equations Eqs.~(\ref{e2})
and~(\ref{e1}), can be transformed to an autonomous system of
differential equations by means of the transformations
$q=2GM/r+GM_{U}/r$, $\mu =3\alpha r^{2}U$, $p=3\alpha r^{2}P$, and $
\theta =\ln r$ where $\mu $ and $p$ are the ``reduced'' dark
radiation and pressure, respectively. With the use of the new
variables, Eqs.~(\ref{e2}) and~(\ref{e1}) become
\begin{equation}
\frac{dq}{d\theta }=\mu -q,  \label{aut1}
\end{equation}
\begin{equation}
\frac{d\mu }{d\theta }=-\frac{\left( 2\mu +p\right) \left[ q+\frac{1}{3}%
\left( \mu +2p\right) \right] }{1-q}-2\frac{dp}{d\theta }+2\mu
-2p. \label{aut2}
\end{equation}

Equations.~(\ref{e2}) and~(\ref{e1}) or, equivalently, Eqs.~(\ref{aut1})
and~(\ref {aut2}), are called the structure equations of the
vacuum on the brane \cite {Ha03}. In order to close the system of
Eqs.~(\ref{aut1}) and~(\ref {aut2}), an ``equation of state''
$p=p\left( \mu \right) $, relating the reduced dark radiation and
the dark pressure terms, is needed. Once the equation of state is
known, exact vacuum solutions of the gravitational field equations
on the brane can be obtained. The opposite procedure can also be
followed: that is, by specifying the functional form of the metric
tensor, the dark radiation and the dark pressure can be obtained
from the field equations. Therefore, several exact solutions of
the gravitational field equations on the brane can be obtained
\cite{Da00,cfm02,BMD03}.

\begin{center}
\section{Accretion by brane-world black holes}
\end{center}

The braneworld description of our universe entails a large extra
dimension and a fundamental scale of gravity that might be lower
by several orders of magnitude compared to the Planck scale
\cite{RS99a,RS99b}. It is known that the Einstein field equations
in five dimensions admit more general spherically-symmetric black
holes on the brane than four-dimensional general relativity. Hence,
an interesting consequence of the brane-world scenario is in the
nature of spherically-symmetric vacuum solutions to the brane
gravitational field equations, which could represent black holes
with properties quite distinct from those of ordinary black holes
in four dimensions. Such black holes are likely to have very
diverse cosmological and astrophysical signatures. In the present
section, we consider the accretion properties of several brane-
world black holes, which have been obtained by solving the vacuum
gravitational field equations. There are many black-hole-type
solutions on the brane, and in the following, we analyze three
particular examples. In all cases, we assume that the inflowing gas
obeys the polytropic equation of state.

\subsection{The DMPR Brane-world Black Hole}

 The first brane-world black hole we consider is the solution of the vacuum field equations
 obtained by Dadhich, Maartens, Papadopoulos  and Rezania  \cite{Da00}, which represents the
simplest generalization of the Schwarzschild solution of general
relativity. We call this type of brane-world black hole as the DMPR
black hole. For this solution, the metric tensor components are
given by
\begin{equation}
e^{\nu }=e^{-\lambda }=1-\frac{2m}{r}+\frac{Q}{r^{2}},
\end{equation}
where $Q$ is the so-called tidal charge parameter. In the limit $%
Q\rightarrow 0$, we recover the usual general relativistic case.
The metric is asymptotically flat, with $\lim _{r\to
\infty}\exp{(\nu )}=\lim _{r\to \infty}\exp{(\lambda )}=1$. There
are two horizons, which are given by
\begin{equation}
r_{h}^{+,-}=m\pm \sqrt{m^{2}-Q}.
\end{equation}
Both horizons lie inside the Schwarzschild horizon $r_{s}=2m$,
$0\leq r_{h}^{-}\leq r_{h}^{+}\leq r_{s}$. In the brane-world
models, there is also the possibility of a negative $Q<0$, which
leads to only one horizon $r_{h+}$ lying outside the Schwarzschild
horizon,
\begin{equation}
r_{h+}=m+\sqrt{m^{2}+Q}>r_{s}.
\end{equation}
In this case, the horizon has a greater area than its general
relativistic counterpart, so that bulk effects act to increase the
entropy and decrease the temperature and to strengthen the
gravitational field outside the black hole.

For the matter inflowing onto the black hole, we adopt again the
equations of state given by Eqs. (\ref{eqs}), (\ref{eqs1}) and
(\ref{eqs2}). For the velocities at the sonic point, we immediately
obtain
\begin{equation}
u_{c}^{2}=\frac{1}{2}\left(
\frac{m}{r_{c}}-\frac{Q}{r_{c}^{2}}\right) ,
\end{equation}
and
\begin{equation}
V_{c}^{2}=\frac{u_{c}^{2}}{1-3u_{c}^{2}-Q/r_{c}^{2}},
\end{equation}
respectively. With the use of Eq. (\ref{velt}), we obtain the
temperature of the gas at the sonic point as
\begin{equation}
T_{pc}=\frac{nu_{c}^{2}}{(n+1)\left[1-\left( 3+n\right)
u_{c}^{2}-Q/r_{c}^{2}\right]}=
\frac{n\left(m/r_{c}-Q/r_{c}^{2}\right)}{%
(n+1)\left[2-(3+n)m/r_{c}+(1+n)Q/r_{c}^{2}\right]}.
\end{equation}

For the adopted metric, the equations describing the steady flow of
the matter onto the black hole are given by
\begin{equation}
\left( \frac{c^{2}}{K}\right) ^{n}T_{p}^{n}ur^{2}=D_{1},
\label{bl1}
\end{equation}
and
\begin{equation}
\left[ 1+\left( n+1\right) T_{p}\right] ^{2}\left( 1-\frac{2m}{r}+\frac{Q}{%
r^{2}}+u^{2}\right) =D_{3},  \label{bl2}
\end{equation}
respectively, with $D_{1}$ and $D_{3}$ being constants of integration.
Evaluating Eqs. (\ref{bl1}) and (\ref{bl2}) at $r=r_{\infty }$,
where $e^{\nu }=e^{-\lambda }\approx 1$ and the velocity is
$u\left( r_{\infty }\right) =u_{\infty }$, gives
\begin{equation}
D_{1}=\left( \frac{c^{2}}{K}\right) ^{n}T_{p\infty }^{n}u_{\infty
}r_{\infty }^{2},
\end{equation}
and
\begin{equation}
D_{3}=\left[ 1+\left( n+1\right) T_{p\infty }\right] ^{2}\left(
1+u_{\infty }^{2}\right) ,
\end{equation}
respectively. Therefore, the two algebraic equations of motion of
the gas in the unknowns $T_{p}$ and $u$ are
\begin{equation}\label{bl3}
T_{p}^{n}ur^{2}=T_{p\infty }^{n}u_{\infty }r_{\infty }^{2},
\end{equation}
and
\begin{equation}\label{bl4}
\left[ 1+\left( n+1\right) T_{p}\right] ^{2}\left( 1-\frac{2m}{r}+\frac{Q}{%
r^{2}}+u^{2}\right) = \left[ 1+\left( n+1\right) T_{p\infty
}\right] ^{2}\left( 1+u_{\infty }^{2}\right) .
\end{equation}

By assuming that $T_{p}<<1$ and $u_{\infty }<<1$, Eq. (\ref{bl4})
immediately gives
\begin{equation}
u^{2}(r)\approx \frac{2m}{r}-\frac{Q}{r^{2}}.
\end{equation}
By substituting this expression of the
velocity into Eq.~(\ref{bl3})  we obtain the temperature profile of the gas as
\begin{equation}
T_{p}(r)\approx T_{p\infty }\frac{\left( u_{\infty }r_{\infty
}^{2}\right) ^{1/n}}{r^{1/n}\left(2m\right)^{1/2n}\left(
r-Q/2m\right) ^{1/2n}}.
\end{equation}
The density of the gas varies as a function of the radial distance
as
\begin{equation} \rho \left( r\right) \approx \frac{\rho
_{0}}{r\sqrt{r-Q/2m}},
\end{equation}
where we have denoted
\begin{equation}
\rho _{0}=\left( \frac{c^{2}}{K}\right) ^{n}T_{p\infty
}^{n}\frac{u_{\infty }r_{\infty }^{2}}{\sqrt{2m}}.
\end{equation}
At the sonic point, the speed of sound $a_{c}$ is given by
\begin{equation}
a_{c}=\frac{cu_{c}}{\sqrt{1-\left( 3+n\right)
u_{c}^{2}-Q/r_{c}^{2}}}=
\frac{c%
\sqrt{m/r_{c}-Q/r_{c}^{2}}}{\sqrt{2-(3+n)m/r_{c}+(1+n)Q/r_{c}^{2}}}.
\end{equation}

From an observational point of view, it is important to estimate the
physical properties of the gas at the event horizon. At the event
horizon, the gas is moving at the speed of light,
$u_h=u\left(r_h\right)\approx 1$. By taking $r=r_{h}^{+}$, we
obtain for the surface temperature of the black hole
\begin{equation}
T_{p}\left( r_{h}^{+}\right) \approx T_{p\infty }\frac{\left(
u_{\infty }r_{\infty }^{2}\right) ^{1/n}}{\left( m+\sqrt{m^{2}\pm
Q}\right)^{2/n} }.
\end{equation}
The density of the gas traveling through the event horizon is
given by
\begin{equation}
\rho \left( r_{h}^{+}\right) \approx \left( \frac{c^{2}}{K}\right)
^{n}\frac{T_{p\infty }^{n}u_{\infty }r_{\infty }^{2}}{\left(
m+\sqrt{m^{2}\pm Q}\right) ^{2}}.
\end{equation}

On a qualitative level, the DMPR brane-world black hole displays
all the typical accretion properties of the standard general
relativistic black holes. In particular, the speed of the gas at
the event horizon equals the speed of light, and this value is
independent of the mass of the black hole. The temperature and the
density of the gas at the event horizon are modified due to the
presence of the tidal charge $Q$.

\subsection{The CFM Brane-world Black Hole}

Two families of analytic solutions in the brane-world model, which are
parameterized by the ADM mass and the PPN parameters $\beta $ and
$\gamma $ and which reduce to the Schwarzschild black hole for
$\beta =1$, have been found by Casadio, Fabbri, and Mazzacurati
\cite{cfm02}. We call the corresponding brane-world black holes as the
CFM black holes.

The first class of solutions is given by
\begin{equation}
e^{\nu }=1-\frac{2m}{r},
\end{equation}
and
\begin{equation}
e^{\lambda }=\frac{1-\frac{3m}{r}}{\left( 1-\frac{2m}{r}\right) \left[ 1-%
\frac{3m}{2r}\left( 1+\frac{4}{9}\eta \right) \right] },
\end{equation}
respectively, where $\eta =\gamma -1=2\left( \beta -1\right) $. As
in the Schwarzschild case, the event horizon is located at
$r=r_h=2m$. The solution is asymptotically flat; that is,
$\lim_{r\rightarrow \infty }e^{\nu }=e^{\nu _{\infty
}}=\lim_{r\rightarrow \infty }e^{\lambda }=e^{\lambda _{\infty }}=1$.  Then, Eqs. (%
\ref{uc}), (\ref{vc}), and {\ref{ac}) give the velocities at the
sonic point as
\begin{equation}
u_{c}^{2}=\frac{m}{2r_{c}}\frac{1-\frac{3m}{2r_{c}}\left(
1+\frac{4}{9}\eta \right) }{1-\frac{3m}{2r_{c}}},
\end{equation}
\begin{equation}
V_{c}^{2}=\frac{m}{2r_{c}-3m}.
\end{equation}
and
\begin{equation}
\frac{a_c^2}{c^2}=\frac{m}{2r_c-(n+3)m}.
\end{equation}
The temperature of the gas at the sonic point is
\begin{equation}
T_{pc}=\frac{nm}{\left( n+1\right) \left[ 2r_{c}-\left( n+3\right) m\right] }%
=\frac{n}{n+1}\frac{a_c^2}{c^2}.
\end{equation}
The velocity, temperature, and density profiles of the inflowing
gas are represented by
\begin{equation}
u\left( r\right) \approx \sqrt{\frac{2m}{r}}\sqrt{\frac{1-\frac{3m}{2r}%
\left( 1+\frac{4}{9}\eta \right) }{1-\frac{3m}{r}}},
\end{equation}
\begin{equation}
T_{p}(r)\approx T_{p\infty }\frac{\left( u_{\infty }r_{\infty
}^{2}\right) ^{1/n}}{\left( 2m\right) ^{1/2n}r^{3/2n}},
\end{equation}
and
\begin{equation}
\rho \left( r\right) \approx \left( \frac{c^{2}}{K}\right) ^{n}\frac{%
T_{p\infty }^{n}u_{\infty }r_{\infty }^{2}}{\sqrt{2m}r^{3/2}},
\end{equation}
respectively. At the event horizon,
\begin{equation}
u_{h}\approx \sqrt{1-\frac{4}{3}\eta },T_{ph}\approx T_{p\infty }\frac{%
\left( u_{\infty }r_{\infty }^{2}\right) ^{1/n}}{\left( 2m\right) ^{2/n}}%
,
\end{equation}
\begin{equation}
\rho _{h}\approx \left( \frac{c^{2}}{K}\right)
^{n}\frac{T_{p\infty }^{n}u_{\infty }r_{\infty }^{2}}{\left(
2m\right) ^{2}}.
\end{equation}

The accretion properties of this brane-world black hole are very
similar to the standard general relativistic ones. The temperature
and the density distribution of the gas at the event horizon are the
same as in the Schwarzschild case. The speed of the gas at the
event horizon is very slightly modified by a term proportional to
the small parameter $\eta $.

The second class of solutions corresponding to brane-world black
holes \cite{cfm02} has the metric tensor components
given by
\begin{equation}\label{cfm021}
e^{\nu }=\left[ \frac{\eta +\sqrt{1-\frac{2m}{r}\left( 1+\eta \right) }}{%
1+\eta }\right] ^{2},
\end{equation}
and
\begin{equation}\label{cfm022}
e^{\lambda }=\left[ 1-\frac{2m}{r}\left( 1+\eta \right) \right]
^{-1},
\end{equation}
respectively.  The metric is asymptotically flat. In the case
$\eta >0$, the only singularity in the metric is at
$r=r_{0}=2m\left( 1+\eta \right) $, where all the curvature
invariants are regular. $r=r_{0}$ is a turning point
for all physical curves. For $\eta <0$, the metric is singular at $%
r=r_{h}=2m/\left( 1-\eta \right) $ and at $r_{0}$, with
$r_{h}>r_{0}$. $r_{h} $ defines the event horizon.

For this brane-world black hole, the velocities at the sonic point
are given by
\begin{equation}
u_{c}^{2}=\frac{m\left( 1+\eta \right) }{2r_{c}}\frac{\sqrt{1-\frac{2m}{r_{c}%
}\left( 1+\eta \right) }}{\eta +\sqrt{1-\frac{2m}{r_{c}}\left(
1+\eta \right) }},
\end{equation}
\begin{equation}
V_{c}^{2}=\frac{m\left( 1+\eta \right) }{2r_{c}\left[ 1+\eta \sqrt{1-\frac{2m%
}{r_{c}}\left( 1+\eta \right) }\right] -3m\left( 1+\eta \right) },
\end{equation}
\begin{equation}
a_{c}=c\sqrt{\frac{m\left( 1+\eta \right) }{2r_{c}\left[ 1+\eta \sqrt{1-%
\frac{2m}{r_{c}}\left( 1+\eta \right) }\right] -(n+3)m\left( 1+\eta \right) }%
}.
\end{equation}
The sonic temperature is $T_{pc}=na_c^2/(n+1)c^2$. The velocity, temperature and density profiles are
\begin{equation}
u(r)\approx \left( 1+\eta \right) \sqrt{1-\left[ \frac{\eta +\sqrt{1-\frac{2m%
}{r}\left( 1+\eta \right) }}{1+\eta }\right] ^{2}}
\frac{\sqrt{1-\frac{2m}{%
r_{}}\left( 1+\eta \right) }}{\eta +\sqrt{1-\frac{2m}{r_{}}\left(
1+\eta \right) }},
\end{equation}
\begin{equation}
T_{p}(r)\approx T_{p\infty }\frac{\left( u_{\infty }r_{\infty
}^{2}\right) ^{1/n}\left( 1+\eta \right) ^{1/n}}{r^{2/n}\left\{
\left( 1+\eta \right) ^{2}-\left[ \eta +\sqrt{1-\frac{2m}{r}\left(
1+\eta \right) }\right] ^{2}\right\} ^{1/2n}},
\end{equation}
\begin{equation}
\rho \left( r\right) \approx \left( \frac{c^{2}}{K}\right)
^{n}T_{p\infty
}^{n}\frac{\left( u_{\infty }r_{\infty }^{2}\right) \left( 1+\eta \right) }{%
r^{2}\sqrt{\left( 1+\eta \right) ^{2}-\left[ \eta +\sqrt{1-\frac{2m}{r}%
\left( 1+\eta \right) }\right] ^{2}}}.
\end{equation}

In order to estimate the physical quantities at the surface of the
black hole, we have to consider separately the cases $\eta >0$ and
$\eta <0$. For $\eta >0$ at the singular point
$r_{0}$, we obtain
\begin{equation}
u_{0}=u\left( r_{0}\right) \approx 0,\eta >0
\end{equation}
\begin{equation}
T_{p0}=T_{p}\left( r_{0}\right) = T_{p\infty }\frac{\left(
u_{\infty }r_{\infty }^{2}\right) ^{1/n}}{\left( 2m\right)
^{2/n}\left( 1+\eta \right) ^{1/n}\left( 1+2\eta \right)
^{1/2n}},\eta >0,
\end{equation}
\begin{equation}
\rho _{0}=\rho \left( r_{0}\right) \approx \left(
\frac{c^{2}}{K}\right) ^{n}T_{p\infty }^{n}\frac{u_{\infty
}r_{\infty }^{2}}{\left( 2m\right)^2 \left( 1+\eta \right) \left(
1+2\eta \right) ^{1/2}},\eta >0.
\end{equation}
For this brane-world black hole, the four-velocity of the inflowing
gas at the singular point $r=r_0$ is zero. At this point, all the
curvature invariants are regular \cite{cfm02}.  However, the
temperature and the density distribution of the gas at the singularity
are very similar to the Schwarzschild black hole case.

For $\eta <0$, the physical parameters at the event horizon have
the values
\begin{equation}
u_{h}\rightarrow \infty ,\eta <0,
\end{equation}
\begin{equation}
T_{ph}\approx T_{p\infty }\frac{\left( u_{\infty }r_{\infty
}^{2}\right) ^{1/n}\left( 1+\eta \right) ^{2/n}}{\left( 2m\right)
^{2/n}},\eta <0,
\end{equation}
\begin{equation}
\rho _{h}\approx \left( \frac{c^{2}}{K}\right) ^{n}T_{p\infty }^{n}\frac{%
u_{\infty }r_{\infty }^{2}\left( 1+\eta \right) ^{2}}{\left( 2m\right) ^{2}}%
,\eta <0.
\end{equation}
For this brane-world black hole model, the four-velocity of the gas
diverges at the event horizon. A typical trajectory approaching
and possibly entering the horizon is such that the physical radius
always decreases (as in the Schwarzschild case) and hits the
singularity at $r_h>0$ \cite{cfm02}. However, the divergence of
the gas velocity at $r_h$ may indicate that the case $\eta <0$ is
unphysical.

\subsection{The BMD Brane-world Black Hole}

Several classes of brane world black hole solutions have been
obtained by Bronnikov, Melnikov, and Dehnen \cite{BMD03} (for
short the BMD black holes). In the following, we analyze the
accretion properties of a particular class of these models, with a
metric given by
\begin{equation}
e^{\nu }=\left( 1-\frac{2m}{r}\right) ^{2/s},e^{\lambda }=\left( 1-\frac{2m}{%
r}\right) ^{-2},
\end{equation}
where $s\in N$. The metric is asymptotically flat, and at
$r=r_h=2m$, these solutions have a double horizon.

For the sonic velocities and temperature, we obtain
\begin{equation}
u_{c}^{2}=\frac{1}{s}\frac{m}{r_{c}}\left( 1-\frac{2m}{rc}\right)
,
\end{equation}
\begin{equation}
V_{c}^{2}=\frac{m\left( 1-\frac{2m}{r_{c}}\right) }{sr_{c}+m\left( 1-\frac{2m%
}{r_{c}}\right) },
\end{equation}
\begin{equation}
T_{pc}=\frac{n}{n+1}\frac{a_{c}^{2}}{c^{2}}=\frac{n}{n+1}\frac{m\left( 1-\frac{2m}{r_{c}}%
\right) }{sr_{c}+(1-n)m\left( 1-\frac{2m}{r_{c}}\right) }.
\end{equation}

The velocity, temperature, and density profiles are given by
\begin{equation}
u\left( r\right) \approx \sqrt{1-\left( 1-\frac{2m}{r}\right)
^{2/s}}\left( 1-\frac{2m}{r}\right) ^{1-1/s},
\end{equation}
\begin{equation}
T_{p}\left( r\right) \approx T_{p\infty }\frac{\left( u_{\infty
}r_{\infty
}^{2}\right) ^{1/n}}{r^{2/n}\left[ 1-\left( 1-\frac{2m}{r}\right) ^{2/s}%
\right] ^{1/2n}},
\end{equation}
\begin{equation}
\rho (r)\approx \left( \frac{c^{2}}{K}\right) ^{n}T_{p\infty }^{n}\frac{%
u_{\infty }r_{\infty }^{2}}{r^{2}\left[ 1-\left( 1-\frac{2m}{r}\right) ^{2/s}%
\right] ^{1/2}}.
\end{equation}
For the physical quantities at the event horizon $r=r_h=2m$, we
obtain
\begin{equation}
u_{h}\approx 1,s=1,
\end{equation}
\begin{equation}
u_{h}\approx 0,s>1,
\end{equation}
and
\begin{equation}
T_{ph}\approx T_{p\infty }\frac{\left( u_{\infty }r_{\infty
}^{2}\right) ^{1/n}}{\left( 2m\right) ^{2/n}},\rho _{h}\approx \left( \frac{%
c^{2}}{K}\right) ^{n}T_{p\infty }^{n}\frac{u_{\infty }r_{\infty }^{2}}{%
\left( 2m\right) ^{2}}.
\end{equation}

For these classes of brane-world black holes, the temperature and the
density distribution of the gas at the event horizon is identical
to the Schwarzschild case. The behavior of the four-velocity at
$r_h$ depends on the value of $s$. For $s=1$, the gas
four-velocity tends to the speed of light while for $s>1$, it
tends to zero.

\begin{center}
\section{Discussions and final remarks}
\end{center}

In the present paper, we have considered the basic physical
properties of matter forming a thin accretion disc in the
space-time metric of brane-world black holes. The basic
physical parameters of the inflowing gas-the temperature, density
and velocity profiles-have been explicitly obtained for several
classes of brane-world black holes and for several values of the
parameters characterizing the vacuum solution of the generalized
field equations in the brane-world models. All the astrophysical
quantities related to the observable properties of the accretion
processes can be obtained from the black hole metric.

Due to the differences in the space-time structure, the brane-
world black holes present some important differences with respect
to the accretion properties, as compared to the standard general
relativistic Schwarzschild case. Therefore, the study of
accretion processes by compact objects is a powerful indicator of
their physical nature. Since the temperature and the density
distributions, as well as the location  of the sonic points and
the velocity distributions, in the case of the brane-world black
holes are different from those for the standard general
relativistic case, the astrophysical determination of these
physical quantities could discriminate, at least in principle,
between the different brane-world models and give some constrains
on the existence of extra dimensions.

\section*{Acknowledgments}

This work was supported by the General Research Fund grant number HKU 701808P of the government of the
Hong Kong Special Administrative Region.

\label{lastpage}

\end{document}